\documentclass[11pt]{article}
\usepackage{amssymb,amsfonts,amsmath}
\usepackage{graphicx}
\usepackage{fancyhdr}
\usepackage{eucal}
\usepackage[english]{babel}
\usepackage[usenames, dvipsnames]{color}
\usepackage[perpage]{footmisc}
\usepackage[round, sort, numbers, authoryear]{natbib}
\usepackage{ifthen}
\usepackage{ifpdf}
\usepackage{epstopdf}
\makeatletter
\def\@cite#1#2{{#1\if@tempswa , #2\fi}}
\def\@biblabel#1{}
\makeatother

\newcommand{\be}{\begin{equation}}
\newcommand{\en}{\end{equation}}
\def\pt{{\partial}}
\def\om{{\omega}}
\def\kap{{\kappa}}
\def\bkap{{\boldsymbol\kappa}}
\def\d{{\rm d}}
\def\uv{\mathbf{u}}
\def\vv{\mathbf{v}}
\def\hv{\widetilde{\mathbf{v}}}
\def\bv{\mathbf{b}}

\def\ev{\mathbf{e}}
\def\Bv{\mathbf{B}}

\def\Bz{{B_{0z}}}
\def\cd{{\mathcal{D}}}
\def\cl{{\mathcal{L}}}
\def\cq{{\mathcal{Q}}}
\def\cm{{\bf{\mathcal{M}}}}
\def\rt{^{\rm T}}
\def\mat{}

\begin{document}

\begin{center}
{\bf\Large Optimal Energy Growth in Current Sheets}

\vspace{0.5cm}

David MacTaggart \& Peter Stewart

School of Mathematics \& Statistics

University of  Glasgow

david.mactaggart@glasgow.ac.uk
\end{center}

%%%%%%%%%%%%%%%%%%%%%%%%%%%%%%%%%%%%%%%%%%%%%%%%%%%
%% Authors Names
%
%\author{\inits{D.M.}\fnm{David}~\lnm{MacTaggart}$^{1}$\orcid{0000-0003-2297-9312}}
%\author{\inits{P.S.}\fnm{Peter}~\lnm{Stewart}$^{1}$\orcid{0000-0002-0971-8057}}
%
%%\author{\inits{D.M.}\fnm{David}~\lnm{MacTaggart}$^{1}$}
%%\author{\inits{P.S.}\fnm{Peter}~\lnm{Stewart}$^{1}$}
%
%%\author{D.~\surname{MacTaggart}$^{1}$\sep
% %       P.~\surname{Stewart}$^{1}$  
% %      }
%
%   \institute{$^{1}$ School of Mathematics \& Statistics, University of Glasgow, Glasgow, UK, G12 8SQ\\
%                     email: \url{david.mactaggart@glasgow.ac.uk}\\ 
%             }
%%%%%%%%%%%%%%%%%%%%%%%%%%%%%%%%%%%%%%%%%%%%%%%%%%%%
%%% Runningheads
%%
%\runningauthor{D. MacTaggart, P. Stewart}
%\runningtitle{Optimal Energy Growth in Current Sheets}

%%%%%%%%%%%%%%%%%%%%%%%%%%%%%%%%%%%%%%%%%%%%%%%%%%%
%% Affilations 
%% id shold be the same with \author addressref value.
%\address[id={}]{}

%%%%%%%%%%%%%%%%%%%%%%%%%%%%%%%%%%%%%%%%%%%%%%%%%%%
%%% Abstract 
\begin{abstract}
In this paper, we investigate the possibility of transient growth in the linear perturbation of current sheets. The resistive magnetohydrodynamic (MHD) operator for a background field consisting of a current sheet is non-normal, meaning that associated eigenvalues and eigenmodes can be very sensitive to perturbation. In a linear stability analysis of a tearing current sheet, we show that modes that are damped as $t\rightarrow\infty$ can produce transient energy growth, contributing faster growth rates and higher energy attainment (within a fixed finite time) than the unstable tearing mode found from normal-mode analysis. We determine the transient growth for tearing-stable and tearing-unstable regimes and discuss the consequences of our results for processes in the solar atmosphere, such as flares and coronal heating. {Our results have significant potential impact on how fast current sheets can be disrupted. In particular, transient energy growth due to (asymptotically) damped modes may lead to accelerated current sheet thinning and, hence, a faster onset of the plasmoid instability, compared to the rate determined by the tearing mode alone.}
\end{abstract}

%%%%%%%%%%%%%%%%%%%%%%%%%%%%%%%%%%%%%%%%%%%%%%%%%%%
%% Keywords
%
%\keywords{Magnetohydrodynamics $\cdot$ Instabilities $\cdot$ Magnetic Reconnection, Theory}

%-------------------------------------------------

%%%%%%%%%%%%%%%%%%%%%%%%%%%%%%%%%%%%%%%%%%%%%%%%%%%
%% Sections
%
% \section{}%\label{s:?} 
\section{Introduction}
The prototypical instability in resistive magnetohydrodynamics (MHD) is the {tearing instability}. As the name suggests, this instability describes the growth of the ``tearing'' of magnetic field or, to be more precise, the change in the field's magnetic topology. In two dimensions, the tearing instability creates a series of magnetic islands, similar to Kelvin's cats eyes \citep[\emph{e.g.}][]{schindler06}. In three dimensions, the change in magnetic topology can be more complicated \citep[\emph{e.g.}][]{priest14}. 

The standard magnetic field configuration for the tearing instability is the {current sheet}. The name derives from a thin layer of intense (compared to the surrounding environment) current density located in a highly sheared magnetic field. Normally, the magnetic field points in opposite directions on either side of the current sheet, with the width of the current sheet (where the change takes place) being much smaller than the typical length scale of the large-scale system.  

Since the seminal work of  \cite{furth63}, there have been many studies of the tearing instability that consider effects such as different geometries or the inclusion of extra physics \citep[\emph{e.g.}][]{pritchett80,terasawa83,tassi07,tenerani15}. In the context of solar physics, magnetic reconnection (the change of magnetic topology) is a fundamental physical process, so the tearing instability is of great interest in this field. Solar eruptions, ranging from flares to jets to coronal mass ejections (CMEs) are often believed to be triggered by magnetic reconnection \citep[\emph{e.g.}][]{macneice04,dmac14,dmac15}. MHD simulations have demonstrated that fast eruptive behaviour is strongly linked to the tearing of current sheets. 

Although large-scale MHD simulations, such as those cited above, can describe the nonlinear evolution of the tearing instability, they are not so effective when it comes to analysing the onset of instability. The complex geometries, sensitivity to boundary conditions and low (compared to the corona) Lundquist numbers make a detailed analysis of the onset of instability very challenging. Therefore, studies that focus only on the (linear) onset of the instability are still very important.

When studying the onset of the tearing instability, the vast majority of studies have focussed on {normal-mode analysis} \citep[\emph{e.g.}][]{chandra61}. That is, solutions are sought with a time dependence of the form
\begin{equation}
\phi \sim \exp(-{\rm i}\omega t),
\end{equation} 
 where $\phi$ represents a variable of the system, $\omega$ is the frequency and $t$ is time. If $\Im(\omega)>0$, then $\phi$ grows exponentially as $t\rightarrow\infty$. Otherwise, if $\Im(\omega)<0$, then $\phi$ decays exponentially as $t\rightarrow\infty$. The objective of normal-mode analysis is to find the largest value of $\Im(\omega)$, which corresponds to the fastest growing mode. The onset of the instablilty can, therefore, be recast as an eigenvalue problem for eigenvalues $\omega$.
For the tearing instability, there is one eigenvalue such that $\Im(\omega)>0$. Hence, there is only one mode that causes exponential growth in the linearized system and is referred to as the {tearing mode}. It can be shown analytically \citep{furth63,schindler06} that the growth rate of the tearing mode depends on the magnetic Lundquist number $S$ (which we shall define later) in the form $S^{-\alpha}$, where $0<\alpha<1$. For environments such as the solar corona, where the magnetic Lundquist number is ${\rm O}(10^{8})$ and above \citep[\emph{e.g.}][]{hood11}, the tearing mode growth rate is very slow compared to rapidly occuring phenomena like flares. This has led researchers to study the nonlinear tearing instability in order to find faster dynamics. However, it may be the case that a faster onset of the instability can be found in the analysis of the linearized system by including the energy growth ignored by normal-mode analysis.

As mentioned above, eigenvalues describe the behaviour of growth or decay as $t\rightarrow\infty$. In normal-mode analysis, all eigenvalues satisfying $\Im(\omega)<0$ (exponential decay) are ignored. However, modes associated with these rejected eigenvalues can produce {transient growth} which, although it decays exponentially as $t\rightarrow \infty$, can produce significant energy growth within a finite time. If such transient growth is large enough, the growth of the linear system could enter the nonlinear regime much faster than by the growth rate of the tearing mode alone. {Therefore, the transient growth due to the damped modes of the system could lead to current sheet disruption much faster than by the growth rate of the (unstable) tearing mode.}

There has been a lot of interest in the study of transient growth of the linearlized Navier-Stokes equations for shear flows \citep[\emph{e.g.}][]{reddy93a,reddy93b,schmid94,hanifi96}. Mathematically, transient growth corresponds to the {non-normality} of the system. What characterises a non-normal system is the non-orthogonality of eigenmodes. To analyse the non-normal behaviour of such systems, a generalization of the eigenvalue spectrum, known as the {pseudospectrum}, can be used \citep{trefethen05}. \cite{bobra94} use pseudospectra to relate ideal and resistive MHD spectra. They show that the resistive MHD eigenmodes, for sheared background fields, are strongly non-orthogonal and, hence, can exhibit transient growth. The effects of non-normal behaviour in MHD have also been studied in the context of magnetic field generation \citep[\emph{e.g.}][]{farrell99a,farrell99b,livermore06}. In solar physics,  transient energy growth has attracted attention in solar wind applications \citep[\emph{e.g.}][]{camporeale09,camporeale12}. 

The effects of non-normal behaviour have not, to our knowledge, been applied to eruptive behaviour in the corona, which is the focus of this paper. By considering a sheared background magnetic field (a current sheet) we will study the effects of transient behaviour in the cases when the system is (spectrally) stable and unstable to the tearing instability. We illustrate the non-normality of the associated operator using a particular form of the pseudospectrum that is simple to calculate once the eigenvalue spectrum has been obtained. The paper is outlined as follows: the initial model equations and boundary conditions are introduced, the background theory for calculating the optimal energy growth is discussed, the spectra and energy growth envelopes are displayed for several cases, and the paper concludes with a discussion of potential applications and further work.
 
\section{Model Description}
To study the tearing instability, we consider the 2D incompressible MHD equations

\be\label{mhd1}
\rho\left(\frac{\pt\uv}{\pt t}+(\uv\cdot{\bf\nabla})\uv\right) = - {\bf\nabla} p+\mu^{-1}({\bf\nabla}\times\Bv)\times\Bv,
\en

\be\label{mhd2}
\frac{\pt\Bv}{\pt t} = {\bf\nabla}\times(\uv\times\Bv)+\eta{\bf\nabla}^2\Bv,
\en

\be\label{mhd3}
{\bf\nabla}\cdot\Bv = {\bf\nabla}\cdot\uv = 0,
\en
where $\Bv$ is the magnetic field, $\uv$ is the velocity, $\rho$ is the (constant) density, $p$ is the plasma pressure, $\eta$ is the constant magnetic diffusivity and $\mu$ is the magnetic permeability. Although compressible MHD would be a more suitable model for the solar atmosphere, we choose to use incompressible MHD for two reasons. The first reason is simplicity - to illustrate our procedure, incompressible MHD allows for an obvious measure of the disturbance energy. The theory that we shall develop, however, could be extended to compressible MHD and more complicated models. The second reason is that most of the literature on the tearing instability uses incompressible MHD. Therefore, comparison with previous work can be made more directly.

For our background (static) equilibrium,
\be
\quad p_0 = p_0(x), \quad {\Bv}_0 = \Bz(x)\ev_z, \quad \uv_0={\bf 0},
\en
where the zero subscript corresponds to the equilibrium and
\be\label{equil}
p_0(x) + \frac{1}{2\mu} B_{0z}^2(x) = {\rm const.}
\en
Before choosing a particular form for $B_{0z}(x)$, let us linearize the MHD equations. Setting $(\uv,\Bv,p) = (\uv_0,\Bv_0,p_0) + (\uv_1,\Bv_1,p_1)$   results in the linearization

\be\label{lmhd1}
\rho\frac{\pt\uv_1}{\pt t} = -{\bf\nabla} p_1 + \mu^{-1}({\bf\nabla}\times\Bv_1)\times\Bv_0 + \mu^{-1}({\bf\nabla}\times\Bv_0)\times\Bv_1,
\en

\be\label{lmhd2}
\frac{\pt\Bv_1}{\pt t} = {\bf\nabla}\times(\uv_1\times\Bv_0) + \eta{\bf\nabla}^2\Bv_1
\en

\be\label{lmhd3}
{\bf\nabla}\cdot\Bv_1 = {\bf\nabla}\cdot\uv_1 = 0.
\en
Note that we are assuming $\eta\ll1$ which is typical in many solar and astrophysical applications. We therefore ignore the contribution of diffusion on the background equilibrium in Equation \ref{lmhd2}, expecting the dynamics of the instability to to occur on a much shorter time scale than the diffusion time.

We now look for solutions of the form
\be
\uv_1 = [u(x,t),0,u_z(x,t)]\rt {\rm e}^{ikz}, \quad \Bv_1 = [b(x,t),0,b_z(x,t)]\rt {\rm e}^{ikz},
\en
where $k$ is the wavenumber of disturbances in the $z$-direction. Taking the curl of Equation \ref{lmhd1}, we eliminate $p_1$. Using the solenoidal constraints in Equation \ref{lmhd3}, we can eliminate $u_z$ and $b_z$. This leaves:
\be\label{lin1}
\frac{\pt}{\pt t}\left(\frac{\pt^2u}{\pt x^2}-k^2u\right) = \frac{ik\Bz}{\mu\rho}\left(\frac{\pt^2b}{\pt x^2}-k^2b\right) - \frac{ikB''_{0z}}{\mu\rho}b,
\en

\be\label{lin2}
\frac{\pt b}{\pt t} = ik\Bz u + \eta\left(\frac{\pt^2b}{\pt x^2}-k^2b\right),
\en
where a prime denotes differentiation with respect to $x$.
\subsection{Equilibrium}
We choose a classic form for the background magnetic field known as the {Harris sheet}.  The magnetic field of the Harris sheet is given by
\be
\Bz(x) = B_0\tanh\left(\frac{x}{x_0}\right), \quad B''_{0z}(x) = -\frac{B_0}{x_0^2}\frac{2}{\cosh^2(x/x_0)}\tanh\left(\frac{x}{x_0}\right),
\en
where $B_0$ is the maximal field strength and $x_0$ measures the thickness of the current sheet. The equilibrium pressure then comes from Equation \ref{equil} but is not important for our calculations.

\subsection{Non-dimensionalization}
To non-dimensionalize the equations, consider

\be\label{nd}
u = u_0u^*, \quad b = B_0b^*, \quad t = t_0t^*, \quad x=x_0 x^*,
\en
with
\be
t_0 = \frac{x_0}{u_0}, \quad u_0 = \frac{B_0}{\sqrt{\mu\rho}},
\en
where $u_0$ is the Alfv\'en speed. The linearized MHD equations become (after dropping the asterisks)

\be\label{nlmhd1}
\frac{\pt}{\pt t}\left(\frac{\pt^2u}{\pt x^2}-k^2u\right) = {ik\Bz}\left(\frac{\pt^2b}{\pt x^2}-k^2b\right) - {ikB''_{0z}}b,
\en

\be \label{nlmhd2}
\frac{\pt b}{\pt t} = ik\Bz u + S^{-1}\left(\frac{\pt^2b}{\pt x^2}-k^2b\right),
\en
where
\be
S = \frac{x_0u_0}{\eta}
\en
is the (non-dimensional) Lundquist number.

\subsection{Boundary Conditions}
We require that $b\rightarrow 0$ and $u\rightarrow 0$ as $x\rightarrow\pm\infty$. However, since numerical simulations typically range bewteen finite values, we shall approximate the boundary conditions as $b=u=0$ at $x=\pm d$ for some $d>0$. This approach will make possible comparisons to simulations of tearing instabilities more feasible. Also, since the tearing instability develops in a thin boundary layer near $x=0$, a value of $d$ much larger than the width of the boundary layer will result in a good approximation. In the Appendix, we perform one of our subsequent calculations in the half-plane $(x,z)\in[0,\infty)\times(-\infty,\infty)$. Comparing this to the corresponding result from the closed domain reveals that the exact form of the boundary conditions is not of vital importance for the results of this paper. 

\section{Background Theory}
In this section we discuss the background theory for determining the optimal energy growth and how non-normal contributions are included. { Our aim is to solve the full initial value problem, rather than just the eigenvalue problem. However, in order to determine the effects of different modes on energy growth, we recast the initial value problem in terms of a selection of eigenvalues and (corresponding) eigenmodes. By considering the kinetic and magnetic energies, we define a (physically) suitable norm for the system and use this to determine the optimal energy growth.}

\subsection{Operator Equations}
In anticipation of the numerical approach that we shall describe later, we write the linearized MHD equations as a maxtrix-vector system. Equations \ref{nlmhd1} and \ref{nlmhd2} can be written in the form

\be\label{ivp}
\frac{\pt}{\pt t}\cm\vv = \cl\vv,
\en
with

\be\label{mm}
\cm=\left(\begin{array}{cc}
\cd^2-k^2 & 0\\
0&\mathcal{I}	
\end{array}\right), \quad \cl = \left(\begin{array}{cc}
0 & \cl_I\\
ikB_z&\cl_R
\end{array}\right), \quad \vv = \left(\begin{array}{c}
u\\
b
\end{array}\right),
\en
and
\be
\cl_I = ikB_z(\cd^2-k^2) - ikB_z'', \quad \cl_R = S^{-1}(\cd^2-k^2), \quad \cd = \frac{\pt }{\pt x}, 
\en
where $\mathcal{I}$ represents the identity operator. If we consider solutions of the form
\be\label{et}
\vv = \hv\exp(-i\om t), \quad \omega \in \mathbb{C},
\en
we can transform the initial value problem of Equation \ref{ivp} into the generalized eigenvalue problem
\be\label{evalue}
-i\om \cm\hv = \cl\hv. 
\en
Making the assumption of Equation \ref{et} restricts us to examining growth or decay in the limit of $t\rightarrow\infty$ only.  In normal-mode analysis, we would solve Equation \ref{evalue} for the eigenvalue with the largest value of $\Im(\omega)>0$. This approach, however, misses the possibilty of transient growth due to eigenmodes with corresponding eigenvalues satisfying $\Im(\omega)<0$, {\emph{i.e.} damped modes.}

 In our calculations of transient growth, we shall make use of the eigenvalue spectrum calculated from Equation \ref{evalue} and the corresponding eigenmodes.  In practice, however, we shall only need to consider a finite number of eigenmodes since not all eigenfuctions will contribute non-normal behaviour. Therefore,  we restrict ourselves to the space $\mathbb{S}^N$ spanned by the first $N$ {least damped} eigenmodes of $\cm^{-1}\cl$:
\be
\mathbb{S}^{N} = {\rm span}\{\hv_1,\dots,\hv_N\}.
\en
We expand the vector functions $\vv\in\mathbb{S}^N$ in terms of the basis  $\{\hv_1,\dots,\hv_N\}$:
\be\label{kap}
\vv = \sum^N_{n=1}\kap_n(t)\hv_n.
\en
Note that the expansion coefficients $\kappa_n$ are functions of $t$ since we are solving the full initial value problem of Equation \ref{ivp} and not the restricted problem of Equation \ref{evalue}.  We can restate Equation \ref{ivp} in the simple form
\be
\frac{\d \bkap}{\d t} = -i\Lambda\bkap, \quad \Lambda\in\mathbb{C}^{N\times N}, \quad \bkap \in \mathbb{C}^{N},
\en
with
\be
\bkap = [\kappa_1\dots,\kappa_N], \quad {\Lambda} = {\rm diag}[\om_1\dots,\om_N].
\en
The operator $\Lambda$ represents the linear evolution operator, $\cm^{-1}\cl$, projected onto the space $\mathbb{S}^N$.

\subsection{Energy Norm}
In order to complete the transformation of the vector functions $\vv$ to coefficients $\boldsymbol\kap$, we must consider the scalar product and its associated norm. To measure the disturbance energy, we consider the combination of the (nondimensional) disturbance kinetic and magnetic energies
\be
E_V = \frac12\int_V(|\uv|^2+|\bv|^2)\,\d V.
\en
From Equation (\ref{lmhd3}) we have 

\be
\cd u + iku_z = 0, \quad \cd b + ikb_z = 0.
\en
Therefore, by virtue of Parseval's equivalence \citep[\emph{e.g.}][]{tichmarsh48}, we can write

\be\
E_V =  \int_k\frac{1}{2k^2}\int_{-d}^{d}(|\cd u|^2 + k^2|u|^2 + |\cd b|^2 + k^2|b|^2)\, \d x\,\d k .
\en
Following previous works \citep[\emph{e.g.}][]{reddy93a}, we take the energy density $E$ as
\be\label{eden}
E= \frac{1}{2k^2}\int_{-d}^{d}(|\cd u|^2 + k^2|u|^2 + |\cd b|^2 + k^2|b|^2)\, \d x.
\en
Since Equation \ref{eden} provides a sensible measure of the energy for a given $k$, we define the energy norm as
\be\label{norm}
\|\vv\|_{\rm E}^2 = \frac{1}{2k^2}\int_{-d}^{d}(|\cd u|^2 + k^2|u|^2 + |\cd b|^2 + k^2|b|^2)\, \d x
\en
For any $\vv_1$,$\vv_2\in\mathbb{S}^N$, the inner product associated with the above energy norm can be written as
\be\label{inner}
(\vv_1,\vv_2)_{\rm E} = \frac{1}{2k^2}\int_{-d}^{d}\vv_2^{H}\cq\vv_1\,\d x,
\en
where
\be
\cq = \left(\begin{array}{cc}
k^2-\cd^2&0\\
0&k^2-\cd^2
\end{array}\right),
\en
and the superscript $H$ represents the complex-conjugate transpose. The integrands in Equations \ref{norm} and \ref{inner} can be related \emph{via} integration by parts. Equation \ref{inner} can be written as
\be
(\vv_1,\vv_2)_{\rm E} = \frac{1}{2k^2}\int_{-d}^{d}\vv_2^{H}\cq\vv_1\,\d x = \bkap^{H}\mat{Q}\bkap,
\en
where the matrix $\mat{Q}$ has components
\be
Q_{ij} = (\hv_i,\hv_j)_{\rm E} = \frac{1}{2k^2}\int_{-d}^{d}\hv_j^H\cq\hv_i\, \d x.
\en
The matrix $\mat{Q}$ is both Hermitian and positive definite. We can, therefore, factor $\mat{Q}$ according to $\mat{Q}=\mat{F}^H\mat{F}$ \citep[\emph{e.g.}][]{trefethen97}, leading to
\begin{eqnarray}
(\vv_1,\vv_2)_{\rm E} &=& \bkap_2^H\mat{Q}\bkap_1 \\
&=& \bkap_2^H\mat{F}^H\mat{F}\bkap_1 \\
&=& (\mat{F}\bkap_1,\mat{F}\bkap_2)_2.\label{inner_F}
\end{eqnarray}
The associated vector norm satisfies
\be
\|\vv\|_{\rm E} = \|\mat{F}\bkap\|_2,  \quad \vv\in\mathbb{S}^N.
\en
%For matrices $\mat{P}\in\mathbb{C}^{N\times N}$, the energy norm is induced by the vector norm according to
%\begin{eqnarray}
%\|\mat{P}\|_{\rm E} &=& \max_{\vv\ne {\bf 0}}\frac{\|\mat{P}\vv\|_{\rm E}}{\|\vv\|_{\rm E}} \\
%&=& \max_{\vv\ne {\bf 0}}\frac{\|\mat{F}\mat{P}\vv\|_2}{\|\mat{F}\vv\|_2} = \max_{\vv\ne {\bf 0}}\frac{\|\mat{F}\mat{P}\mat{F}^{-1}\mat{F}\vv\|_2}{\|\mat{F}\vv\|_2} \\
%&=& \|\mat{F}\mat{P}\mat{F}^{-1}\|_{2}.
%\end{eqnarray}
This relationship between the energy norm and the $L^2$ norm will be useful for the practical calculation of the optimal energy growth that we shall discuss shortly.

\subsection{Optimal Growth}
The formal solution of the initial value problem \ref{ivp} can be written as
\be
\vv = \exp(\cm^{-1}\cl t)\vv_0, \quad \vv_0 = \vv(0).
\en
Using Equation \ref{kap} we can transform the above result to 
\be
\bkap = \exp(-i{\Lambda}t)\bkap_0, \quad \bkap_0 = \bkap(0).
\en
The optimal transient growth of the disturbance energy is given by the norm of the matrix exponential
\begin{eqnarray}
G(t) \equiv G(t,S,k) &=& \max_{\vv_0\ne\bf{0}}\frac{\|\vv(t)\|_{\rm E}^2}{\|\vv_0\|_{\rm E}^2} \\
&=& \max_{\bkap_0\ne\bf{0}}\frac{\|F\bkap(t)\|_2^2}{\|F\bkap_0\|_2^2} \\
&=& \max_{\bkap_0\ne\bf{0}}\frac{\|F\exp(-i{\Lambda}t)\bkap_0\|_2^2}{\|F\bkap_0\|_2^2} \\
&=& \max_{\bkap_0\ne\bf{0}}\frac{\|F\exp(-i{\Lambda}t)F^{-1}F\bkap_0\|_2^2}{\|F\bkap_0\|_2^2} \label{pogr}\\
&=&\|\mat{F}\exp(-i{\Lambda}t)\mat{F}^{-1}\|_2^2. \label{ogr}
\end{eqnarray}
Equation \ref{ogr} follows from Equation \ref{pogr} {via the definition of an induced norm}.

The curve traced out by $G(t)$ {\it vs.} $t$ represents the maximum possible energy amplification, which for each instant of time is optimized over all possible initial conditions with unit energy norm \citep{schmid94}. The initial disturbance that optimizes the amplification factor can be different for different times. Therefore, $G(t)$ should be thought of as the {envelope} of the energy growth of individual initial conditions with unit energy norm. Henceforth, we shall refer to $G(t)$ as the {optimal energy envelope}.
\section{Numerical Procedure}
In this section we briefly outline the main numerical procedures for the required calculations. Until now, we have presented the theory in terms of the underlying {operators}. Since a practical solution requires a (finite) discretization of the problem, we shall henceforth refer to {matrices} rather than operators and {eigenvectors} rather than eigenmodes. When referring back to an equation containing operators, it will be implicit that we are now considering the discretized version of that equation and, hence, are strictly dealing with finite matrices rather than operators.
\subsection{Discretization for the Eigenvalue Problem}
We follow previous works on non-normal stability by expanding the variables in terms of Chebyshev polynomials. These functions are defined in the interval $[-1,1]$. It is trivial to convert from the problem domain $[-d,d]$ to the Chebyshev domain \emph{via} $y=x/d$, with $y\in[-1,1]$. A function can be approximated on the Chebyshev interval as
\be\label{cheb}
f(y) = \sum_{n=0}^Na_nT_n(y),
\en 
where 
\be
T_n(y) = \cos[n\cos^{-1}(y)]
\en
and the $a_n$ are constants. The unknown variables, $u$ and $b$ in Equation \ref{ivp} are expanded in the form of Equation \ref{cheb}. Derivatives are also expressed in terms of Chebyshev polynomials and make use of standard recurrence relations \citep[\emph{e.g.}][]{as64}. In order to use these recurrence relations, the expanded equations are then required to be satisfied at the Gauss-Lobatto collocation points
\be
y_j=\cos\left(\frac{\pi j}{N}\right).
\en
If we consider the eigenvalue problem of Equation \ref{evalue}, the expansion in terms of Chebyshev polynomials produces a matrix-vector system where the matrices (for the generalized eigenvalue problem) contain spectral differentiation matrices and the vector contains the expansion coefficients $a_n$. 

Boundary conditions are included in rows of one of the matrices of the discretized generalized eigenvalue problem. The corresponding rows in the other matrix are chosen to be a complex multiple of these rows. By choosing a large complex multiple, spurious modes associated with the boundary conditions can be mapped to a part on the complex plane far from the region of interest (far below the eigenvalues near $\Im(\omega)=0)$. { To illustrate this approach, consider the discrete form of Equation \ref{evalue} 
\be
-i\omega M\mathbf{x} = L\mathbf{x}
\en
where $M$ and $L$ are finite matrices and $\mathbf{x}$ represents an eigenvector. We can write

\be
M = \left(\begin{array}{ccc}
T_0(1) & T_1(1) & \cdots \\
T_0''(y_1)-k^2T_0(y_1) & T_1''(y_1)-k^2T_1(y_1) & \cdots \\
\vdots & \vdots & \vdots \\
T_0''(y_{N-1})-k^2T_0(y_{N-1}) & T_1''(y_{N-1})-k^2T_1(y_{N-1}) & \cdots \\
T_0(-1) & T_1(-1) & \cdots \\
\vdots & \vdots & \ddots \end{array}\right),
\en
where we indicate the layout of  the top-left section of the matrix (see the definition of $\cm$ in Equation \ref{mm}). Boundary conditions have been included in the 1st and $N$th rows. The same rows in $L$ are chosen as a complex multiple of the corresponding rows in $M$ \citep{reddy93b}. In this paper, we multiply the rows by $-8000i$. For brevity, we do not display the full matrix of the discretized problem.
} 

Once the system is fully discretized, the generalized eigenvalue problem can be solved by standard methods. In this paper, we perform the calculations in MATLAB.

\subsection{Optimal Quantities}
\subsubsection{Energy Growth}
To calculate the optimal energy growth, we make use of singular value decomposition (SVD). Writing $A=\mat{F}\exp(-i{\Lambda}t)\mat{F}^{-1}$, we can decompose this matrix as 
\be
AV=\Sigma U,
\en
where $U$ and $V$ are unitary matrices and $\Sigma$ is a matrix containing the singular values, ordered by size. It can be shown that $\|A\|_2=\sigma_1$, where $\sigma_1$ is the largest singular value of $A$ \citep[\emph{e.g.}][]{trefethen97}. \emph{Via} Equation \ref{ogr}, we use this property to determine the optimal energy growth. Again, we use MATLAB to calculate the SVD.

\subsubsection{Optimal Disturbances}\label{opt}
In order to determine the initial disturbance that will create the maximum possible amplification at a given time $t_0$, we can make further use of the SVD. Let $\mat{A}=\mat{F}\exp(-i{\Lambda}t_0)\mat{F}^{-1}$. If $\sigma_1$ is the largest singular value of $\mat{A}$ then, as described above,
\be
\sigma_1=\|\mat{F}\exp(-i{\Lambda}t_0)\mat{F}^{-1}\|_2=\|\exp(-i{\Lambda}t_0)\|_{\rm E}.
\en
If we perform a decomposition, as before, and now focus only on the column vectors of $\mat{U}$ and $\mat{V}$ corresponding to $\sigma_1$, we obtain
\be
\mat{A}{\vv_1}=\sigma_1\uv_1.
\en
The effect of $\mat{A}$ on an input vector $\vv_1$ results in an output vector $\uv_1$ stretched by a factor of $\sigma_1$. That is, $\vv_1$ represents an initial condition that will be amplified by a factor $\sigma_1$ due to the mapping $\mat{F}\exp(-i{\Lambda}t_0)\mat{F}^{-1}$, where $t_0$ is the time when the amplification is reached \citep[\emph{e.g.}][]{schmid94}. On the subspace $\mathbb{S}^N$, the optimal initial disturbance can be expressed as
\be
\bkap_1=\mat{F}^{-1}\vv_1.
\en

\section{Spectra and Perturbed Matrices}
\begin{figure}[h!]
	\centering
	
		{\includegraphics[width=9.5cm]{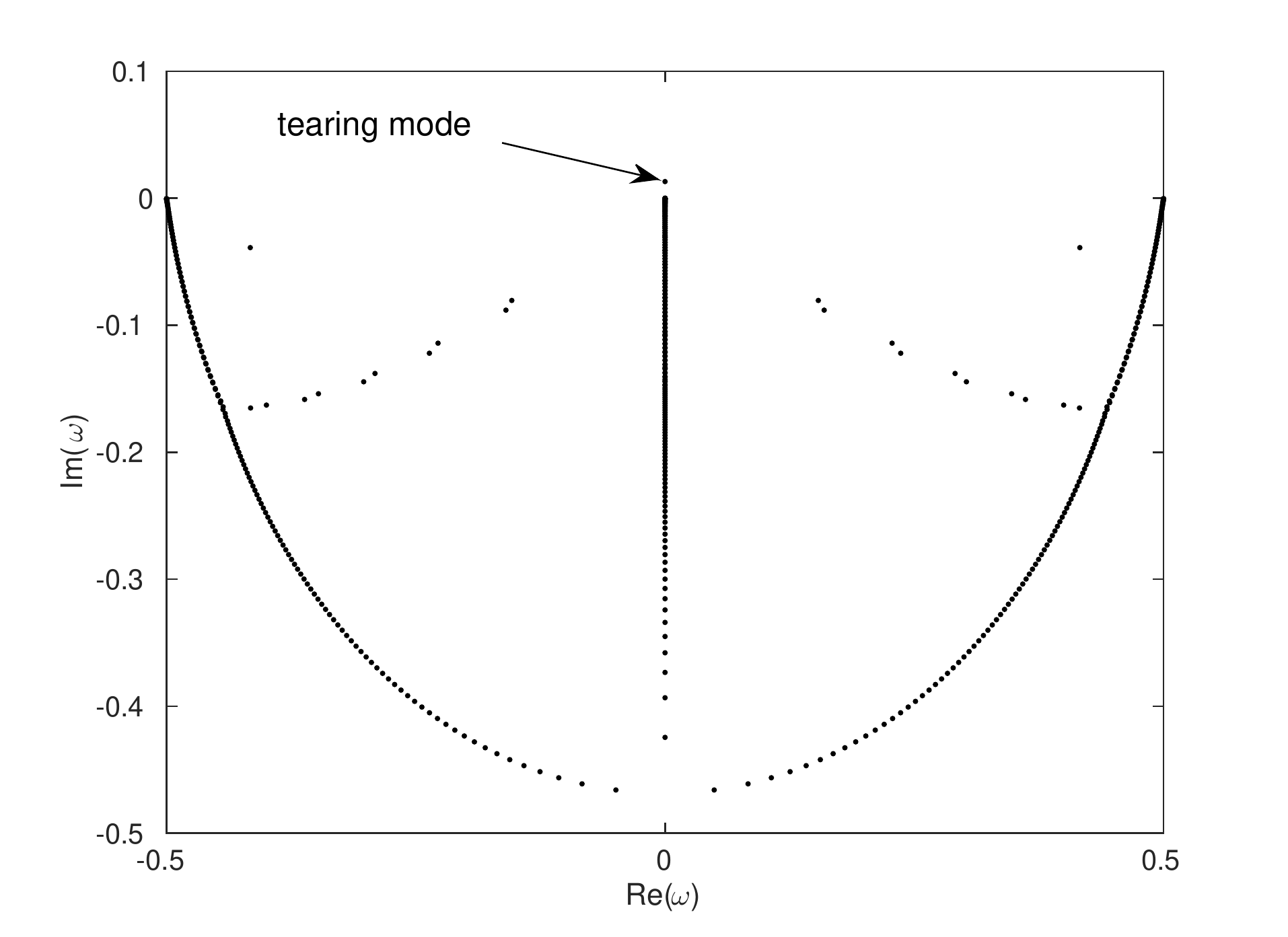}}
 \centerline{     
      \hspace{0.4525 \textwidth}  {(a)}
         \hfill}
		{\includegraphics[width=9.5cm]{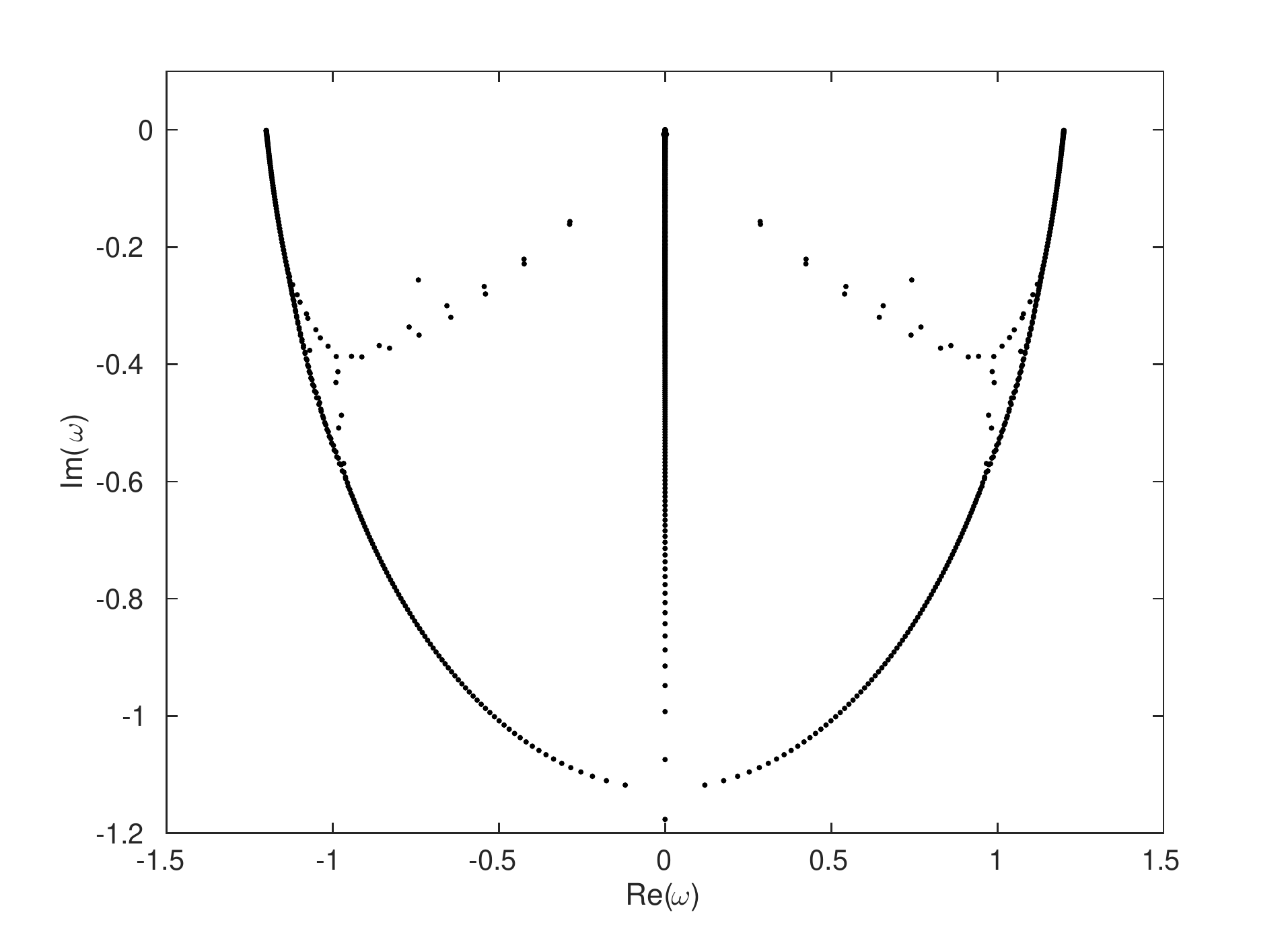}}
 \centerline{     
      \hspace{0.4525 \textwidth}  {(b)}
         \hfill}

 \caption{Spectra for the discrete generalized eigenvalue problem, Equation \ref{evalue}, for $S=1000$ and (a) $k=0.5$ and (b) $k=1.2$. In (a), the unique eigenvalue corresponding to the tearing mode is highlighted.}
\label{spectra}
\end{figure}

In this section we present some of the results from solving the generalized eigenvalue problem of Equation \ref{evalue}. To be more precise, we solve the discretized version of Equation \ref{evalue} subject to the numerical scheme outlined in the previous section. Throughout the rest of the paper, unless specified otherwise, we will set $d=10$. Let us consider $S=1000$ and examine the spectra for the cases $k=0.5$ and $k=1.2$. Figure \ref{spectra} displays these two spectra.

For the tearing problem set up in this paper, it can be shown analytically that the equilibrium can only become tearing-unstable for $0<k<1$. The spectrum in Figure \ref{spectra}a is for $k=0.5$ and the system is, therefore, linearly unstable to the tearing instability. As can be seen from this spectrum, there is only one unstable eigenvalue, labelled as corresponding to the tearing mode. This eigenvalue is $\approx$0.0131, which is equivalent to the value obtained from a finite difference solution of the same problem (Hood, private communication). The layout of the spectrum is qualitatively similar to other tearing-unstable spectra that have been calculated for similar boundary conditions and background equilibria \citep[\emph{e.g.}][]{goedbloed10}. There is a distinct branching structure that is found in the spectra of many non-normal matrices \citep{reddy93b}.  

In the spectrum for $k=1.2$, in Figure \ref{spectra}b, there are no eigenvalues with $\Im(\omega)>0$. There is still, however, a branching structure similar to the previous spectrum. The branch points of the spectra indicate the non-normal behaviour of this resistive MHD problem. This means that eigenvectors with eigenvalues satisfying $\Im(\omega)<0$ can contribute transient growth to the amplification of energy. In order to reveal this non-normal behaviour, consider the following description. Let $\mat{A}$ be a matrix from which the eigenvalues of the problem are found, and let $\mat{E}$ be a matrix such that $\|\mat{E}\|_2\le 1$. Consider, also, a small parameter $\epsilon\ll 1$. A complex number, $z$, is in the pseudospectrum of $\mat{A}$, $\sigma_{\epsilon}(\mat{A})$, if $z$ is in the spectrum of $\mat{A}+\epsilon\mat{E}$ (a similar statement can be made for finite operators). For a normal matrix, points $z\in\sigma_{\epsilon}$ can differ from corresponding points in the spectrum of $\mat{A}$  by ${\rm O}(\epsilon)$, i.e. by the size of the perturbation \citep[\emph{e.g.}][]{trefethen05}. 

For a non-normal matrix, however, the difference can be much larger. Instead of the eigenvalues of $A+\epsilon E$ differing from those of $A$ by, at most, ${\rm O}(\epsilon)$, they can differ by ${\rm O}(1)$. Such behaviour is particularly present at the branch points of spectra.

\begin{figure}[h!]
	\centering
	
		{\includegraphics[width=9.5cm]{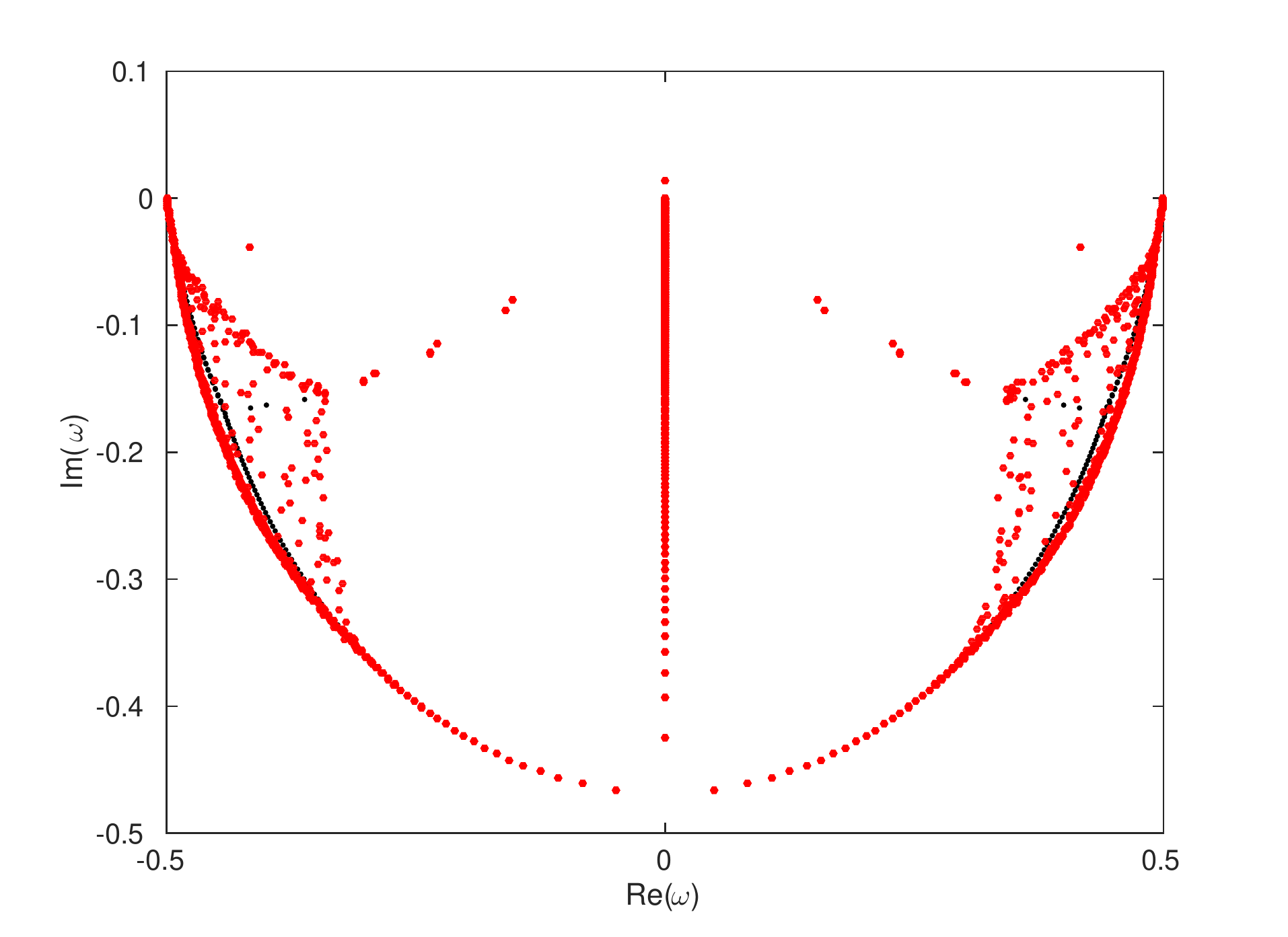}}
 \centerline{     
      \hspace{0.4525 \textwidth}  {(a)}
         \hfill}
		{\includegraphics[width=9.5cm]{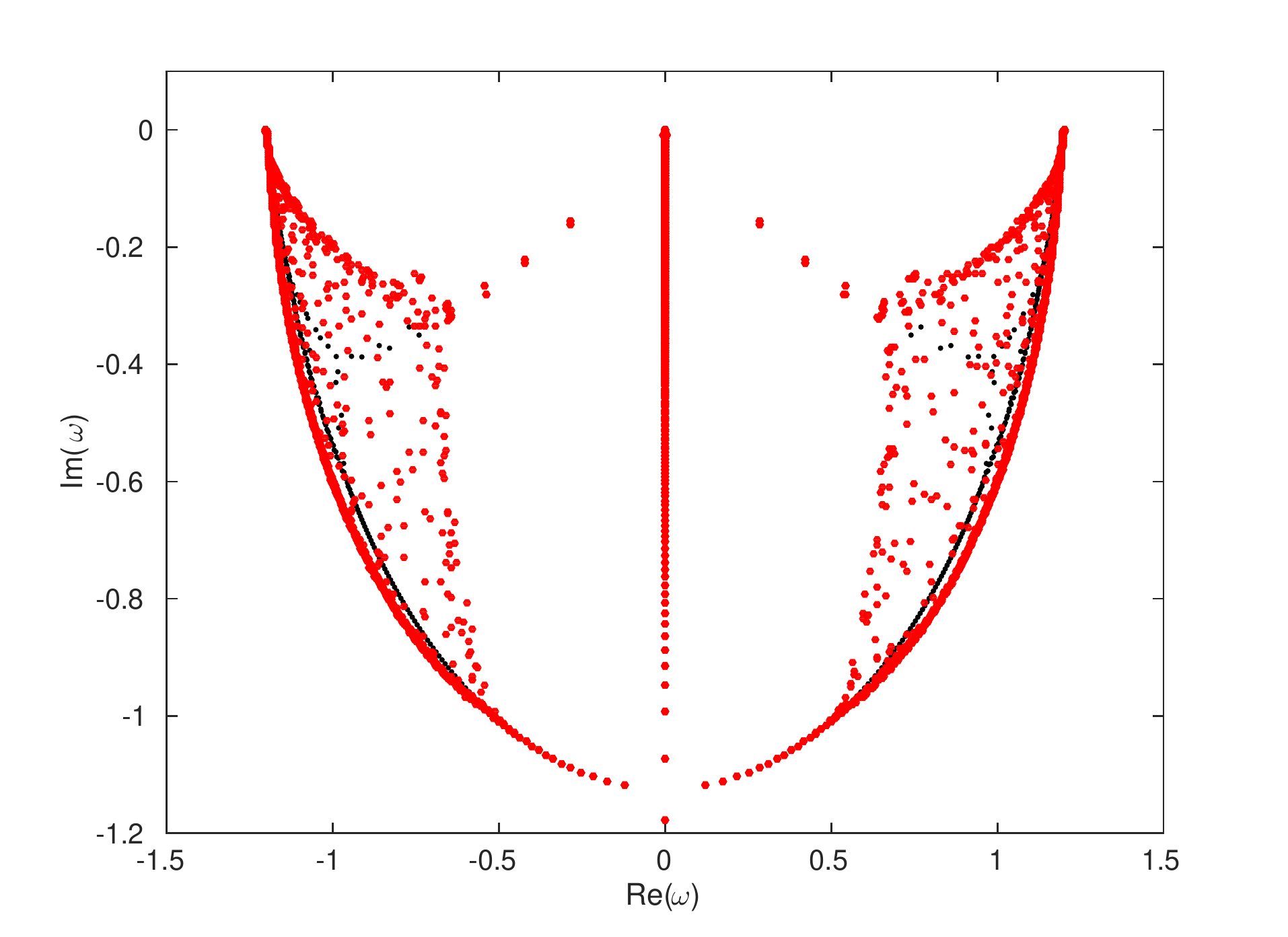}}
 \centerline{     
      \hspace{0.4525 \textwidth}  {(b)}
         \hfill}

 \caption{Same spectra as in Figure \ref{spectra} but now with  eigenvalues of the perturbed matrices included and shown in red.}
\label{per}
\end{figure}

If $\mat{A}$ represents the unperturbed matrix of the spectra displayed in Figure \ref{spectra}, Figure \ref{per} displays the spectra of $\mat{A}+\epsilon\mat{E}$ (for $k=0.5,1.2$) where $\epsilon={\rm O}(10^{-6})$ and the entries of $\mat{E}$ are random and taken from a normal distribution. The eigenvalues of $\mat{A}+\epsilon\mat{E}$, for six different random matrices $\mat{E}$, are shown in red.

The pseudospectrum of $\mat{A}$ would be the subset of the complex plane given by
\be\label{pseudospectrum}
\sigma_{\epsilon} = \bigcup_{\|\mat{E}\|_2\le 1}\sigma(\mat{A}+\epsilon\mat{E}).
\en
As demonstrated in Figure \ref{per}, however, only a few matrices $\mat{E}$ are required to reveal the non-normal character of the matrix $\mat{A}$. 

There are several equivalent definitions of pseudospectra \citep{trefethen05}. The definition we have presented here gives the simplest and most practical demonstration of non-normal behaviour. For our current purposes, this version of the pseudospectrum will suffice.

Looking at the eigenvalues of the perturbed matrix, there are two main features that emerge. The first is that for large parts of the spectra, the eigenvalues of $A+\epsilon E$ differ from the eigenvalues $A$ by ${\rm O}(\epsilon)$, indicating normal behaviour. The second feature is that near the branch points of the spectra, the difference is now much larger. For both spectra displayed, a perturbation of ${\rm O}(10^{-6})$ produces a difference of ${\rm O}(10^{-1})$ between the eigenvalues of the matrices $A$ and $A+\epsilon E$ at the branch locations. This jump of five orders of magnitude is a clear signal of non-normality and, hence, the possibility of significant transient growth. Estimating the pseudospectrum of Equation \ref{pseudospectrum} with just a few random matrices $\mat{E}$ is the recommended approach for determining if the system in question is non-normal since it is easily determined from the spectrum which we use for determining the optimal transient growth. Plotting the pseudospectrum estimate, as done in Figure \ref{per}, also reveals what eigenvectors will produce non-normal effects and then should, therefore, be included in the subspace $\mathbb{S}^N$. 

\section{Optimal Energy Growth}

\subsection{Spectrally-Stable $k$}
As stated previously, the onset of the tearing instability, for the present setup, occurs only for $0<k<1$ in normal-mode analysis. However, as demonstrated in the previous section, the system is non-normal and allows for the possibility of transient growth, even for $k>1$. To get an overview of the optimal energy growth for spectrally stable $k$, we calculate $\max_{t\ge 0}G(t)$ for different $k$. For the calculation of $G(t)$, we only consider contributions from eigenvalues with $-1.4<\Im(\omega)<0$. This will mean that for different values of $k$, different numbers of eigenvalues (and therefore eigenvectors) will be used in the calculations. However, this range captures most of the effects of the non-normality, as suggested by the pseudospectra, and does not disguise the main results. The values of $\max_{t\ge 0}G(t)$ for a range of $k>1$ and for the cases $S=100$ and $S=1000$ are displayed in  Table \ref{table}.

\begin{figure}[h]	
\centering
		{\includegraphics[width=9.5cm]{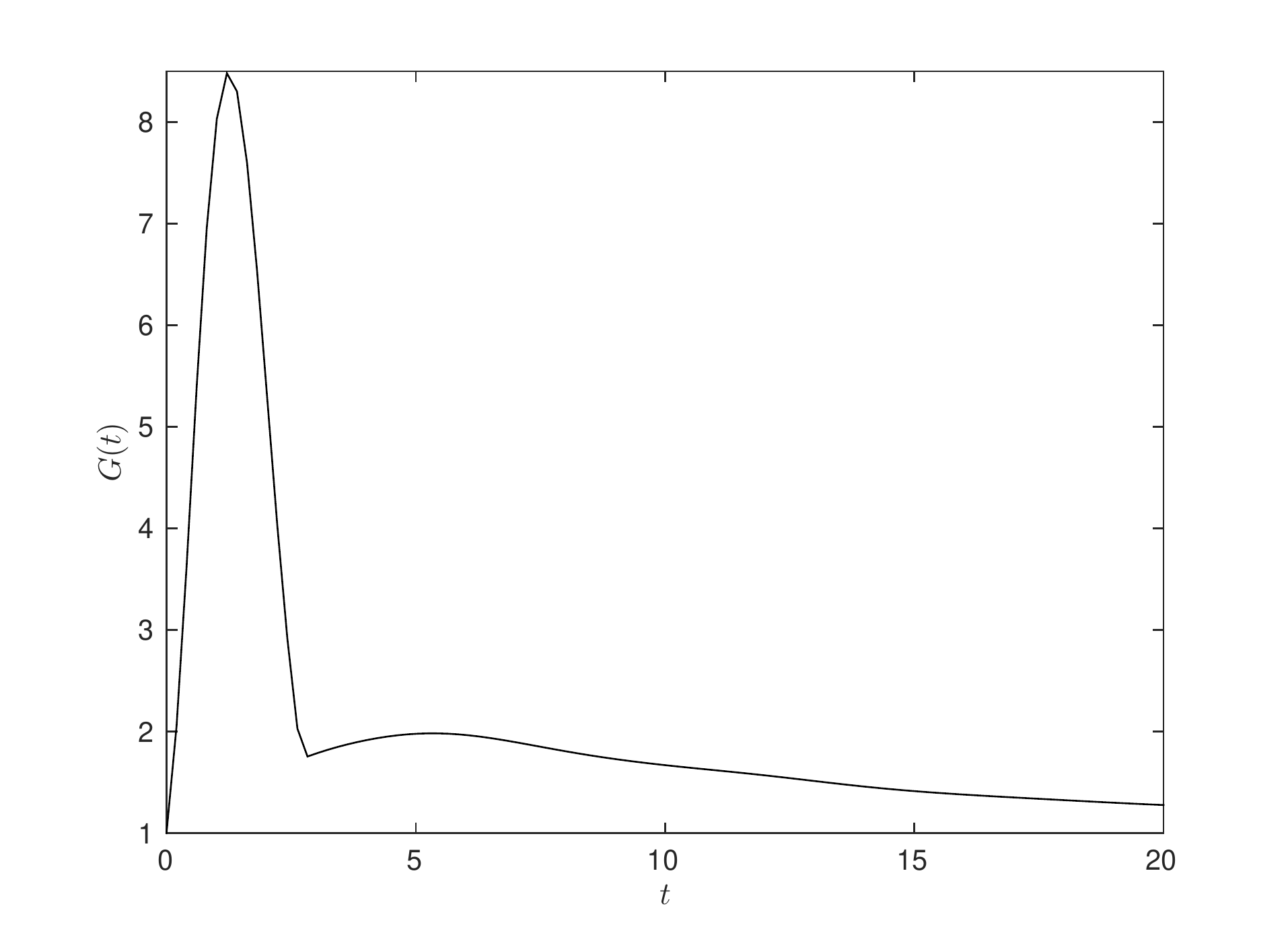}}
 \caption{An example of an optimal energy envelope $G(t)$ for $S=1000$ and $k=1.1$.}
\label{gt_1}
\end{figure}

\begin{table}[h]
  \begin{center}
\def~{\hphantom{0}}
 \caption{Maxima of $G(t)$ in time for $k>1$ and magnetic Lundquist numbers $S=100$, 1000.}
  \label{table}
  \begin{tabular}{lcc}
\hline
      $k$  & $\max G(t,S=100)$   &   $\max G(t, S=1000)$ \\[3pt]
\hline
       1.1   & 1.6 & 8.48 \\
       1.2   & 1.51 & 10.86\\
       1.3  & 1.41 & 11.79  \\
       1.4   & 1.35 & 11.42 \\
       1.5 & 1.29 & 11.57  \\
\hline
\end{tabular}
 
  \end{center}
\end{table}
For $S=100$, the optimal energy growth is small and does not even double in size for the values of $k$ displayed. This result is important as the magnetic Lundquist number for many simulations can be of ${\rm O}(100)$. Therefore, any transient growth would not be noticed. Moving up to $S=1000$, the optimal energy growth can increase by an order of magnitude. In the solar corona, where $S\approx {\rm O}(10^8)$ and higher, it is therefore possible that transient growth for spectrally stable $k$ could become large enough to excite the nonlinear phase of the tearing instability. An example of a $G(t)$ envelope for $k=1.1$, $S=1000$ is shown in Figure \ref{gt_1}.

{In light of the results of Table \ref{table}, we may ask how the transient energy growth can increase with increasing $S$? One way to answer this question is to consider a simple upper bound for the energy growth. For an initial value problem, suppose that $\omega_{\rm I}$ is the imaginary part of the least damped eigenvalue of $\Lambda$. It then follows that
\begin{eqnarray}
\exp(\omega_{\rm I}t) &\le& \|\exp(-i\Lambda t)\|_{\rm E} \\
&\le& \|F\|_2\|F^{-1}\|_2\exp(\omega_{\rm I}t)\\
&\le&\kappa(F)\exp(\omega_{\rm I}t),\label{bound}
\end{eqnarray}
where $\kappa(F)= \|F\|_2\|F^{-1}\|_2$ is the standard notation for the condition number of the matrix $F$ (not to be confused with $\kappa$ from Equation \ref{kap}). If $\kappa(F)=1$ in Equation \ref{bound}, we have equality and the energy bound is determined by the least damped eigenvalue alone. If, however, $\kappa(F) \gg 1$, then there is the potential for substantially larger energy growth at early times, even though it may be that $\omega_{\rm I} <0$. For the tearing-stable case studied above, $\omega_I\approx 0$ and so the energy bound is given by $\kappa(F)$. Table \ref{table2} shows how the condition number varies for some values of $S$ when $k=1.1$.

\begin{table}[h]
  \begin{center}
\def~{\hphantom{0}}
 \caption{$S$ \emph{vs.} $\kappa(F)$ for $k=1.1$.}
  \label{table2}
  \begin{tabular}{ccccccc}
\hline
      $S$  & 10 & 50 & 100 & 500 & 1000 & 5000 \\
\hline
      $\kappa(F)$  & 20 & 690 & 1.6$\times 10^4$ & 2$\times10^8$ & 3$\times 10^8$ & 3.8$\times 10^8$\\
      
\hline
\end{tabular}
 \end{center}
\end{table}
Clearly, using $\kappa(F)$ as an upper bound for the energy is too loose for practical considerations. However, the purpose of displaying these results is to convey the following: as $S$ increases and, hence, the diffusion term in the induction equation is multiplied by a smaller coefficient $S^{-1}$, it may reasonably be expected that the energy bound tends to an ideal MHD limit, where the onset of instability is governed entirely by eigenvalues. However, the opposite is true, allowing for (non-normal) transient effects to play a significant role. As $S$ increases, the eigenvectors (related to $F$ \emph{via} the inner product in Equation \ref{inner_F}) become more ill-conditioned, as discussed in \cite{bobra94}.

Stricter bounds (both upper and lower) for the energy growth can be determined using pseudospectral theory \citep{trefethen05}. However, such considerations go beyond the scope of the present paper and will be considered in future work.  
}

\subsection{Spectrally-Unstable $k$}
For $0<k<1$, a normal-mode analysis would produce the eigenvalue with the highest positive value of $\Im(\omega)$, which would represent the growth rate of the linearly unstable system. For the tearing instability, the growth rate behaves as $S^{-\alpha}$ for $0<\alpha<1$, which, for coronal values, is very slow. {For a discussion the various values of $\alpha$, determined by eigenvalue analysis in different regimes, the interested reader is directed to \cite{tenerani16}.}

\begin{figure}[h]	
\centering
		{\includegraphics[width=9.5cm]{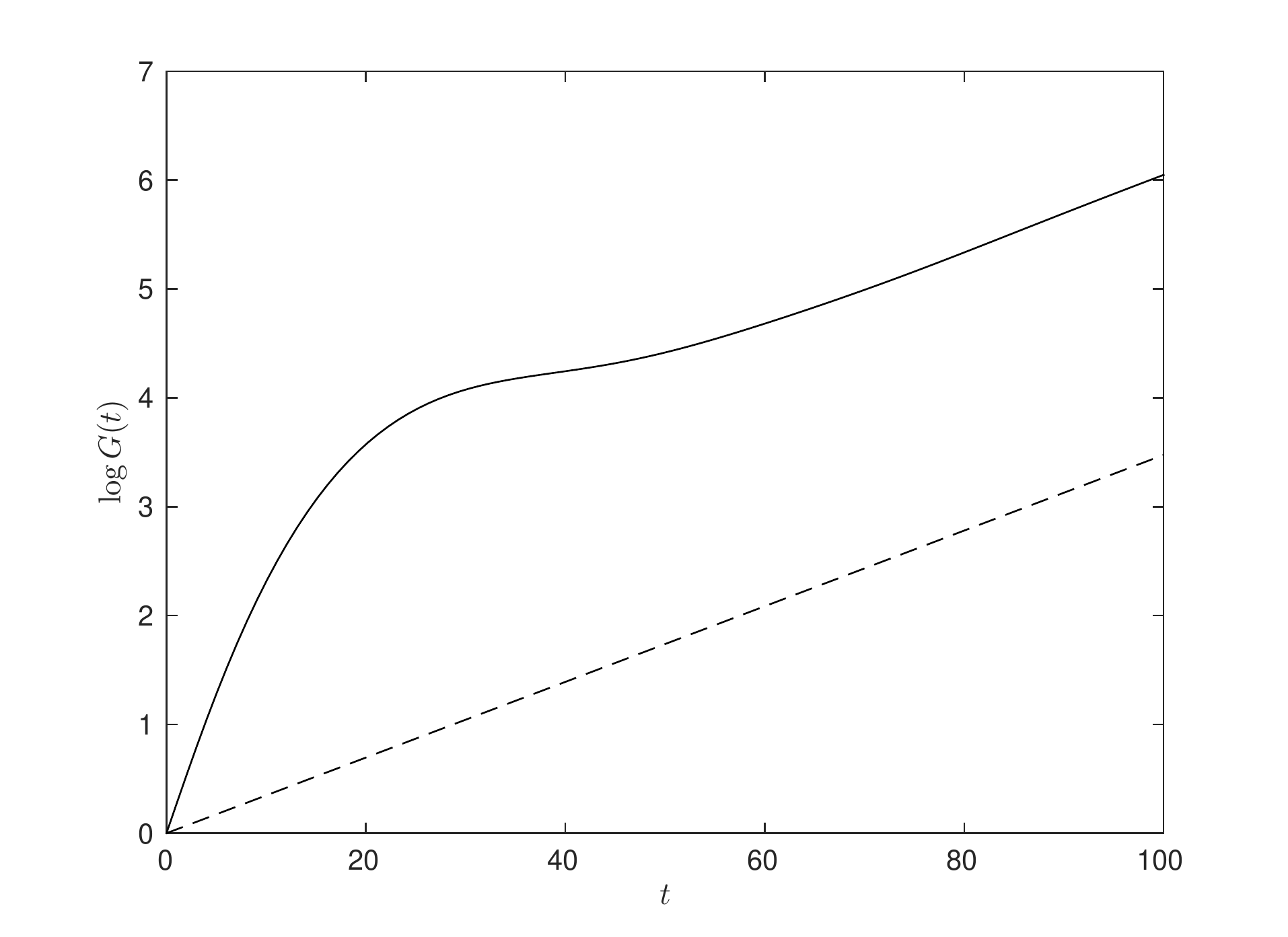}}
 \caption{Optimal energy growth curves $G(t)$ for $S=1000$ and $k=0.2$. Key: $\Im(\omega)>-0.6$ (solid), $\Im(\omega)>0$ (dash).} 
\label{gt_2}
\end{figure}

Since normal-mode analysis ignores any energy growth that decays as $t\rightarrow 0$, the possibility of faster energy growth due to transient effects is often neglected. To demonstrate the possible effect of transients on the growth rate, Figure \ref{gt_2} displays the optimal energy growth envelopes for two cases: the optimal energy growth due to the tearing mode alone and the optimal energy growth due to the combination of the tearing mode and spectrally stable eigenvectors. This example is calculated for $S=1000$ and $k=0.2$ and when transient effects are included, we consider eigenvectors with corresponding eigenvalues with imaginary parts bounded below by $\Im(\omega)=-0.6$.  

The dashed curve represents the optimal energy growth using only the tearing mode. This envelope could be produced if we performed a normal-mode analysis. Comparing this curve to the case where other eigenvectors are included in the calculation reveals very interesting behaviour. By  $t\approx20$, $G(t)$ from the solid curve increases to $\approx30$ (note that the Figure displays $\log G(t)$), compared to that of the dashed curve which only increases to $\approx1$. Including the effects of transient growth has resulted in an optimal energy growth that proceeds much more rapidly, at short times, compared the contribution from the linearly unstable mode alone. By $t\approx 40$, the solid curve begins to plateau and the growth rate is now less than the dashed curve. This is due to the initial transients decaying and having less effect on the energy growth. From $t\approx 60$ and beyond, both curves become parallel. This behaviour is to be expected as the contribution from the unstable mode dominates as $t\rightarrow\infty$. It is clear from Figure \ref{gt_2} that including the effects of the transients can increase the optimal energy growth substantially.

As mentioned before, the curves of $G(t)$ are envelopes of the optimal energy growth and so, in practice, they may not be reached if the initial perturbation is not optimal. However, what Figure \ref{gt_2} reveals is that even if the optimal energy growth is not attained, the gap between the envelopes for growth with and without transient effects can be large. Hence, even a non-optimal perturbation can produce fast energy growth that could amplify the energy to an order of magnitude (or more) greater than that predicted by normal-mode analysis, within a given time. 

\subsection{Optimal Distubances}

\begin{figure}[h]
	\centering
	
		{\includegraphics[width=9.5cm]{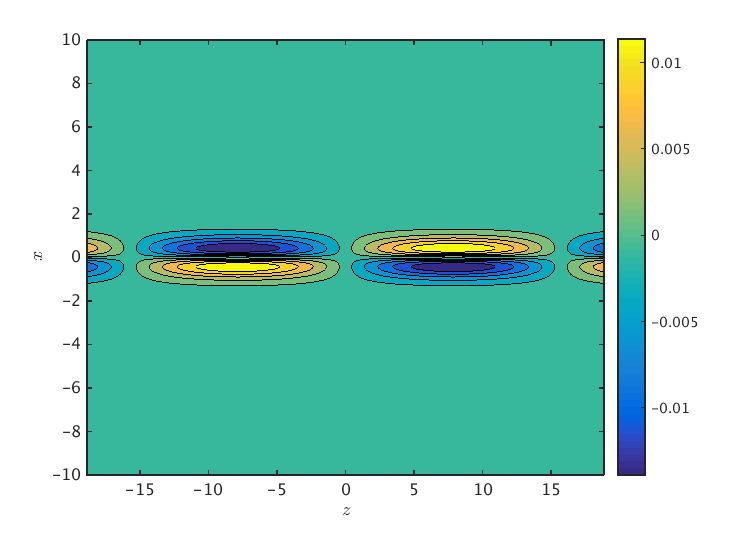}}
  \centerline{     
      \hspace{0.4525 \textwidth}  {(a)}
         \hfill}
		{\includegraphics[width=9.5cm]{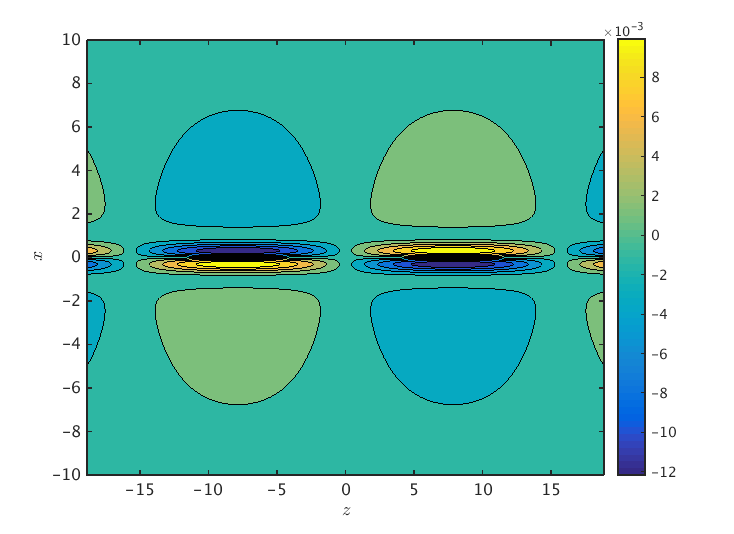}}
  \centerline{     
      \hspace{0.4525 \textwidth}  {(b)}
         \hfill}
 \caption{Optimal initial $u_x$ for (a) $t=30$, (b) $t=40$.}
\label{ini}
\end{figure}

The optimal energy envelopes described in the last section represent, at every point in time, the energy amplification optimized over all initial conditions with unit energy norm. As described in Section \ref{opt}, we can determine the optimal perturbation from the same analysis used to calculate $G(t)$. That is, for a given time, we can determine the initial perturbation that produces the optimal energy amplification at that time. To illustrate this, Figure \ref{ini} shows the optimal initial values for the $x$-component of the velocity at times $t=30$, 40 for the case $S=1000$, $k=0.2$.

The other components of $\uv$ and $\bv$ at $t=0$ can also be found. For brevity, we omit displaying them here. The purpose of calculating the optimal initial conditions is described in the following section.

%\subsection{Reconnection rates}
%One quantity of interest in studies of the tearing instability is the \emph{reconnection rate}. To find the reconnection rate for the present setup, consider Ohm's law for resistive MHD
%\begin{equation}
%\Ev+\uv\times\Bv=\eta\nabla\times\Bv,
%\end{equation}
%where $\Ev$ is the electric field and all other variables have their ususal meanings.

\section{Discussion}

\subsection{Summary}
In this paper, we have demonstrated that the linear onset of the tearing instability can exhibit large transient energy growth due to the non-normality of the associated resistive MHD operator. This energy amplification is found by solving the full initial value problem rather than just the eigenvalue problem of normal-mode analysis. The latter theory is only concerned with the asymptotic growth of the linear system and ignores transient effects.  From our illustrative examples we have shown that transient energy growth can be amplified much faster than than that determined purely from normal-mode analysis. This behaviour has been demonstrated for both tearing-stable and tearing-unstable values of the wavenumber.

To determine the optimal energy growth, we have made use of the eigenvalues and eigenvectors of the system. By plotting pseudospectra, we reveal that a subset of eigenvectors contributes to transient energy growth. The eigenvectors of this subset have eigenvalues $\omega$ with $\Im(\omega)<0$, which are ignored by normal-mode analysis. 

The optimal energy envelopes that we calculate increase in amplitude with the magnetic Lundquist number $S$. These curves represent the possible energy amplification that can be achieved if the initial condition is optimal. However, even if the initial condition is not optimal, there is still the possibility for energy growth that is much faster than the growth rate determined from normal-mode analysis. This means that transient energy growth could, potentially, trigger the nonlinear phase of the tearing instability much sooner than previously expected. If this is the case, the implications for the tearing instability in solar physics would be substantial. 

\subsection{Solar Applications}

\subsubsection{Coronal Phenomena}
In the solar corona, two important phenomena that are often linked to current sheets and their dissipation are {coronal heating} and {solar eruptions}. For the first of these, the ``nanoflare'' theory suggests that the corona is heated by many ``small'' heating events (or flares) spread throughout the coronal magnetic field \citep{parker88}. The tearing of current sheets, that develop from the complex deformation of magnetic fields, is one possible way that the magnetic field can release its energy as heat. Recent models of the nonlinear development of the MHD kink instability have revealed the development of many small-scale current features that could act as nanoflares \citep[\emph{e.g.}][]{hood16}. Our results support the idea of coronal heating \emph{via} tearing instabilities as perturbations could excite large transient growth which, in turn, could potentially readily generate nanoflares.

For the second phenomenon, current sheets are believed to play an important role at the onset, and subsequent nonlinear development, of solar eruptions. Such current sheets would be manifest in the flares associated with the initiation of CMEs, jets and surges. Simulations of CME-type eruptions often reveal a combination of reconnection above and below the CME, referred to as the {breakout theory} of CMEs. In particular, simulations, both 2D and 3D, demonstrate that tearing reconnection above and below the CME heralds the onset of an eruption \citep[\emph{e.g.}][]{macneice04,dmac14}.  The onset of jets and surges has also been linked to the tearing of current sheets \citep[\emph{e.g.}][]{dmac15}. The onset of jets and eruptions is an important topic, not only for theoretical interest but for space weather applications. Therefore, understanding all aspects (normal and non-normal) of the onset of the tearing instability is vital. 

\subsubsection{Quasi-Singular Current Sheets and the Plasmoid Instability}
Recent work by \cite{pucci14} has highlighted that the aspect ratio of current sheets has a threshold value, after which, equilibrium cannot be reached and the current sheet must reconnect. Various simulations have revealed that a fast tearing instability can develop for large $S$ and have growth rates proportional to $S^{1/4}$ \citep{lourerio07,lapenta08}. Hence, in the limit as $S\rightarrow\infty$, there would be, in the words of \cite{pucci14}, an ``infinitely unstable mode'' which is impossible in ideal MHD. By a simple and clever scaling argument, they show that once the current sheet aspect ratio is ${\rm O}(S^{1/3})$, a laminar current sheet cannot be supported and fast tearing must proceed. 

Although we agree with main conclusion of \cite{pucci14}, we would suggest an alternative path to reaching their result. Their analysis is based entirely on eigenvalues and eigenvectors and so ignores the contribution of any transient growth. As the possible energy amplification of transient growth increases with $S$, a much faster onset of the tearing instability could be found that is due to transient growth. Such transient growth depends on the initial perturbation. Hence, the result of \cite{pucci14} can be thought of as a lower bound, when there are no effects of transient growth. As soon as there are perturbations that can induce transient growth, energy amplification will grow faster, thus exciting the tearing instability faster, as shown in the example in Figure \ref{gt_2}.  

{Further recent work by \cite{comisso16} attempts to describe a general theory of the plasmoid instability, formulated by means of a principle of least time. In their analysis, they find that the scaling relationships for the final aspect ratio, the transition time to rapid onset, the growth rate and the number of plasmoids depend on the size of the initial disturbance amplitude, the rate of current sheet evolution, and the Lundquist number. We agree that the initial conditions are important for the onset of the instability, however, we would suggest that the theory of \cite{comisso16} could be extended to include transient effects like those described in this paper. Using scalings for the tearing mode alone will not give a complete description of the transient phase of the instability.
}

\subsection{Future Work}
This work can proceed in two main directions. The first is to include extra physics (\emph{e.g.} two fluid effects) to study how this would effect transient growth. The second, and perhaps most important, is to use optimal initial perturbations as initial conditions in nonlinear resistive MHD simulations. This task will determine if transient growth can lead to a fast nonlinear phase of the tearing instability or if nonlinear terms saturate the transient growth. It will be particularly interesting to determine if the nonlinear tearing instability can be excited by perturbations with $k>1$, \emph{i.e.} spectrally-stable perturbations. 

{Although we have suggested that our results can extend those of previous studies (such as \cite{pucci14} and \cite{comisso16}) there remains much further work to understand how the damped part of the eigenvalue spectrum perturbs the current sheet and drives reconnection, particularly at very high values of $S$.}

%\section*{Disclosure of Potential Conflicts of Interest}

%%%%%%%%%%%%%%%%%%%%%%%%%%%%%%%%%%%%%%%%%%%%%%%%%%%%%%%%%%%%%%%%%%%%%%%%%%%
%% Appendix
%
 \section*{Appendix}   
Throughout this paper we have performed calculations with boundary conditions $u=b=0$ at $x=\pm d$. This has been done so that our results can be easily compared to other works and to nonlinear simulations which typically use such boundary conditions. Since the tearing instability develops in a boundary layer near $x=0$, the precise nature of the boundary conditions should not play a strong role on the onset of the instability. To illustrate this, we solve the discrete form of Equation \ref{evalue}, with $S=1000$ and $k=0.5$, in the {half-plane} and compare the resulting spectrum to that in Figure \ref{spectra}a. 

\begin{figure}[h]
	\centering
		{\includegraphics[width=9.5cm]{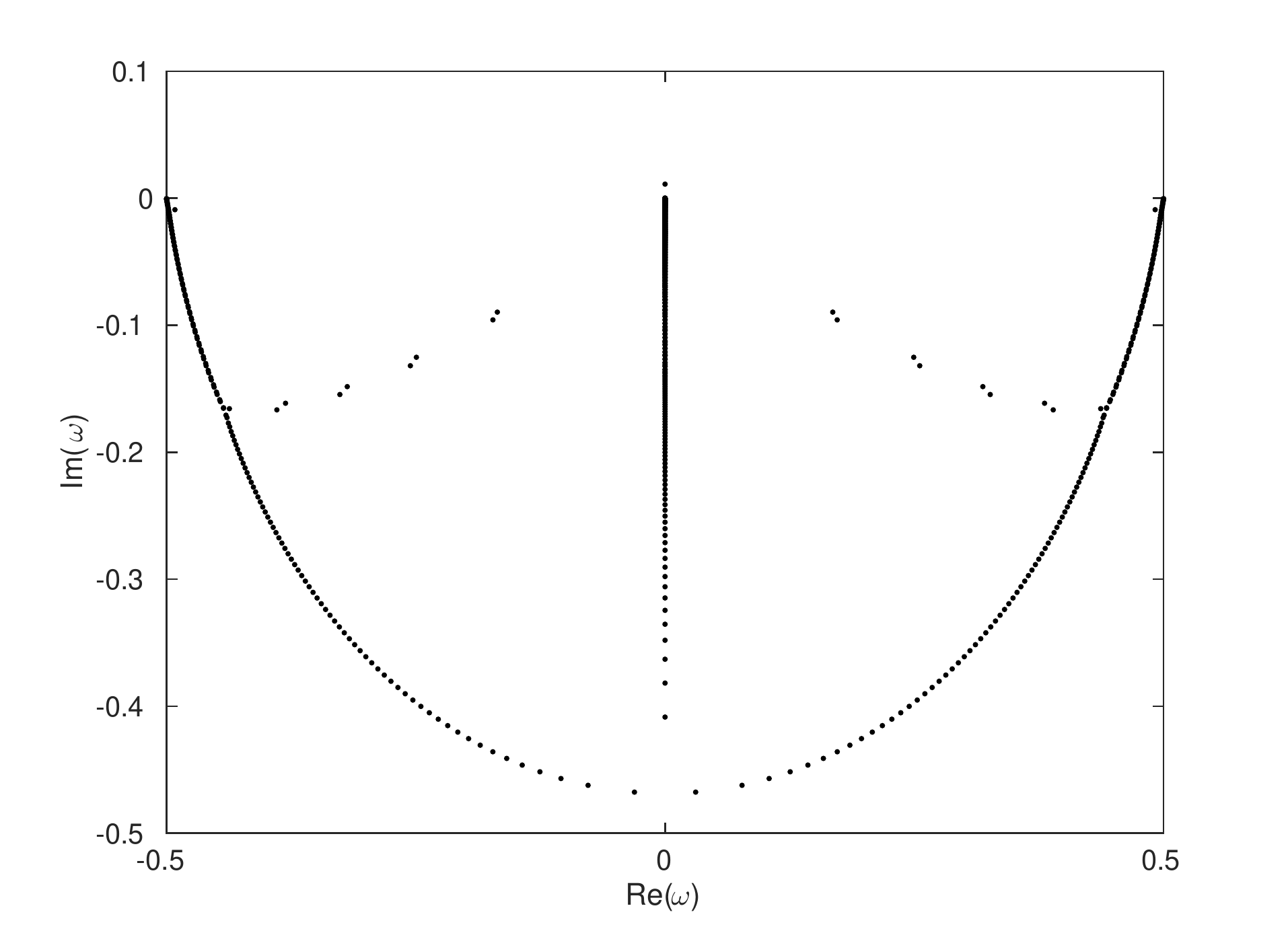}}
  %\vspace{-0.35\textwidth}   % Shift close to the panel top 
  
 \caption{Spectrum of the discrete eigenvalue problem with half-plane boundary conditions for $S=1000$, $k=0.5$.}
\label{app}
\end{figure}

Anticipating a symmetric solution in $\bv$ and an antisymmetric solution in $\uv$ about $x=0$, we set the boundary conditions at $x=0$ to be
\be
u=\frac{{\rm d}b}{{\rm d}x}=0.
\en
As $x\rightarrow\infty$, we set
\be
u=b=0.
\en
In order to represent this boundary numerically, we consider a large domain denoted by $0\le x\le x_{\rm max}$. In order to expand the variables using Chebyshev polynomials, we need to map our coordinates to the domain $-1\le y \le 1$. This is achieved through
\be
x = a\frac{1+y}{b-y},
\en
where
\be
a = \frac{x_{\rm max}x_i}{x_{\rm max}-2x_i} \quad \mbox{and} \quad b=1+\frac{2a}{x_{\rm max}}.
\en
This mapping clusters the grid points near the boundary layer at $x=0$ and places half of the grid points in the region $0\le x\le x_i$ \citep{hanifi96}. In this example, we take $x_{\rm max}$=100 and $x_i=15$. The resulting spectrum is displayed in Figure \ref{app}.

By inspection, the comparison with Figure \ref{spectra}a yields few differences. The eigenvalue corresponding to the tearing mode now has a value $\approx0.0111$, which is still similar to that calculated for the other boundary conditions. Two isolated eigenvalues near $\Im(\omega)=0$ in Figure \ref{spectra}a are now pushed nearer the main branches in Figure \ref{app}. Apart from these minor differences, the spectra calculated from different boundary conditions are very similar. This result suggests that the exact form of boundary conditions, assuming they do not interfere dynamically with the boundary layer at $x=0$, will not radically change the behaviour of the onset of the tearing instability.

%\section{Order of magnitude estimate}
%In this appendix, we show that the growth rate $\sigma\sim O(R^{-3/5})$ using a simple order of magnitude analysis. The approach could be made more rigorous using a matched aspymptotic expansion but we achieve the same result. Let $\sigma$ be the growth rate and $\epsilon$ the thickness of the boundary layer. Then
%\be
%\frac{\partial u}{\partial t} \sim \sigma u, \quad \frac{\partial^2u}{\partial x^2} \sim \epsilon^{-2}u.
%\en
%At the edge of the boundary layer, the background magnetic field can be approximated by
%\be
%B_z(x)\sim B_0\epsilon.
%\en
%%%%%%%%%%%%%%%%%%%%%%%%%%%%%%%%%%%%%%%%%%%%%%%%%%%%%%%%%%%%%%%%%%%%%%%%%%%
%% Acknowledgements
%
% \begin{acks}
%
% \end{acks}

%%% %%%%%%%%%%%%%%%%%%%%%%%%%%%%%%%%%%%%%%%%%%%%%%%%%%%%%%%%%%%
%% Bibliography
%
% Using BibTeX
%
% \bibliographystyle{spr-mp-sola}
% \bibliography{<bib file>}  
%
% Without BibTeX 

\end{document}